\documentclass[prd,tightenlines,nofootinbib,amsfonts,amssymb,amsmath,color,11pt]{article}
\usepackage[utf8]{inputenc}
\usepackage[affil-it]{authblk}
\usepackage{amsmath}
\usepackage{amssymb}
\usepackage{hyperref}
\usepackage{geometry}
\usepackage{color}
\usepackage{graphicx}
\usepackage{epstopdf}
\usepackage[font=small,labelfont=bf]{caption}

\def\be{\begin{equation}} 
\def\ee{\end{equation}}
\def\bea{\begin{eqnarray}}
\def\eea{\end{eqnarray}}

\begin{document}

\title{\bf Anisotropic  cosmological solutions in $R + R^2$ gravity}
\author[1]{Daniel M\"uller \thanks{muller@fis.unb.br}}
\author[2]{Angelo Ricciardone \thanks{angelo.ricciardone@uis.no}}
\author[3,4]{Alexei A. Starobinsky \thanks{alstar@itp.ac.ru}}
\author[4,5]{Aleksey Toporensky \thanks{atopor@rambler.ru}}

\affil[1]{Instituto de F\'{i}sica, Universidade de Bras\'{i}lia, Caixa Postal
04455, 70919-970 Bras\'{i}lia, Brazil}
\affil[2]{Faculty of Science and Technology, University of Stavanger, 4036, Stavanger, Norway}
\affil[3]{L. D. Landau Institute for Theoretical Physics RAS, Moscow 119334, Russian Federation}
\affil[4]{Kazan Federal University, Kazan 420008, Republic of Tatarstan, Russian Federation}
\affil[5]{Sternberg Astronomical Institute, Moscow University, Moscow 119991, Russian Federation}

\date{\today}
 \maketitle
\abstract{\small
 In this paper we investigate the past evolution of an anisotropic Bianchi I universe in $R+R^2$ gravity. Using the dynamical system approach we show that there exists a new two-parameters set of solutions that includes both an isotropic ``false radiation" solution and an anisotropic generalized Kasner solution, which is stable. We derive the analytic behaviour of the shear from a specific property of $f(R)$ gravity and the analytic asymptotic form of the Ricci scalar when approaching the initial singularity. Finally we numerically checked our results.

} 
 \bigskip
\bigskip
 
\section{Introduction}
\label{sec:Intro}
Inflation is a generic intermediate attractor in the direction of expansion of the universe, and in the case of pure $R^2$ gravity it is an exact attractor~\cite{Starobinsky:1987zz}. However, it is not an attractor in the opposite direction in time. Thus, if we are interested in the most generic behavior before  
inflation, more general anisotropic and inhomogeneous solutions should be  considered. We know from General Relativity (GR) that already anisotropic homogeneous solutions
help us much in understanding the structure of a generic space-like curvature singularity. Thus, a natural question is to investigate anisotropic
solutions in the $R+R^2$ gravity, too.\\
In light of the latest Cosmic Microwave Background constraints by PLANCK~\cite{Ade:2015lrj}, the pioneer inflationary model based on the modified $R+R^2$ gravity
(with small one-loop corrections from quantum fields) ~\cite{Starobinsky:1980te} represents one of the most favourable models. It lies among the simplest ones from all viable inflationary models since it contains only one free ajustable parameter taken from observations. Also it provides a graceful exit from inflation and a natural mechanism for creation of known matter after its end, which is actually the same as that used to generate scalar and tensor
perturbations during inflation. 
This theory can be read as a particular form of $f(R)$-theories of gravity which, in turn, is a limiting
case of scalar-tensor gravity when the Brans-Dicke parameter 
$\omega_{BH}\to 0$, and it contains an additional scalar degree of freedom
(scalar particles, or quasi-particles, in quantum language) compared to GR 
which is purely geometrical. The existence of a scalar degree of freedom 
(an effective scalar field) is needed if we want to generate scalar (matter)
inhomogeneities in the universe from ``vacuum" fluctuations of some quantum
field~\cite{Mukhanov:1990me, Lyth:2009zz}. Such generalizations of the familiar Einstein-Hilbert action have been also studied as an explanation for dark energy and late-time acceleration of the universe's expansion~\cite{Capozziello:2002rd, Amendola:2006kh,Nojiri:2006su,Starobinsky:2007hu} and to include quantum behaviour in the gravitational theory~\cite{Stelle:1976gc}.\\ 
As is already very well known, through the Gauss-Bonnet term which in four dimensions is a surface term, the most general theory up to quadratic in curvature terms is of the type $R+R^2+C_{abcd}C^{abcd}$, where $C_{abcd}$ is the Weyl tensor. The investigations of this type of models began with  \cite{weyl1918gravitation,buchdahl1962gravitational,ruzmaikina1970quadratic,Gurovich:1979xg}. After them, many authors  have been analyzed the cosmological evolutions of such a model \cite{tomita1978anisotropic,Muller:1987hp,Berkin:1991nb,Barrow:2005qv,vitenti2006numerical,muller2006starobinsky,cotsakis2008slice,Barrow:2009gx,muller2011homogeneous,de2012bianchi,Muller:2012wx,MULLER:2014jaa,0264-9381-27-22-225013,PhysRevD.77.103523,PhysRevD.75.123515}. A particular attention is given to the asymptotic behavior in \cite{Cotsakis:2007un, Cotsakis:1997ck, Miritzis:2003eu, Miritzis:2007yn}. The addition of a Ricci square term creates a richer set of solutions, and in particular, in~\cite{Starobinsky:1987zz,Muller:1989rp, Barrow:2006xb}, has been shown that the cosmic no-hair theorems no longer hold.\\
Quadratic theory like $R+R^2$ gravity, is a particular case of the more general quadratic type, and has higher order time
derivatives in the equations of motion and this leads to appearance of solutions which have no analogs in GR. One of such solutions corresponds to the scale factor $a \sim \sqrt{t}$ behavior, 
that coincides with a radiation dominated solution in GR. In quadratic gravity, this solution represents a vacuum solution which is also a stable past attractor for Bianchi I model and probably for all Bianchi models~\cite{Barrow:2006xb}. 
The other solution, being an analog of GR solution with matter with equation
of state $p=(\gamma-1)\rho$, is $a \sim t^{4/3 \gamma}$ (instead of usual $a \sim t^{2/3 \gamma}$ in GR),
and it can be naively  considered as a solution which would describe the last stages of a collapsing universe
when quadratic terms dominate. However, this solution appears to be a saddle, so
a collapsing universe for a general initial condition ends up with a vacuum ``false radiation'' regime,
in principle not possible in GR.\\
The $a(t)\propto\sqrt{t}$ behavior near singularity in the $R+R^2$ model 
does not mean that the $R^2$ term behaves as radiation generically. This 
behavior is specific only for the purely isotropic case, as it will be 
shown in the present paper (and even in the isotropic case the late-time behavior 
is different: $a(t)\propto t^{2/3}$ modulated by small high-frequency 
oscillations). Neither does it behave as an ideal fluid in the anisotropic
case.

When shear is taken into account the situation becomes more complicated.
Vacuum solutions exist in GR also, and in the simplest case of a flat anisotropic
metrics (that is the case analyzed in the present paper) this is the famous
Kasner solution~\cite{Kasner:1921zz}. On the other hand, studies of cosmological evolution in a
general quadratic gravity (which includes apart from $R^2$ the Weyl tensor square term in the quadratic part of the action) indicate that the isotropic vacuum ``false radiation'' 
still exists and, moreover, it is an attractor~\cite{Barrow:2006xb}. Kasner solution is also a solution in
quadratic gravity. Due to complicated nature of dynamics near the Kasner solution
a generic trajectory could end up in Kasner or isotropic solution depending
on initial conditions~\cite{Toporensky:2016kss}.

So, for these reasons, the stability and the full behavior of quadratic theories of gravity is still subject of 
 investigation and one powerful tool to address these problems is the dynamical system approach~\cite{wainwright2005dynamical} which allows to find exact solutions 
 of the theory through the determination of fixed points and gives a description of the evolution of the system, at least at qualitative level. 

Despite the obvious fact that the general quadratic gravity at the level of the action includes $R+R^2$ gravity
(which can be obtained setting the coefficient before Ricci square term to zero) it is not so at the level
of corresponding equations of motion for the universe model in question. The reason is that the number
of degrees of freedom in a general quadratic gravity is bigger than in $R+R^2$ theory (which, on its turn 
is bigger than  in GR). That is why we can not simply put corresponding constant to zero in 
cosmological equations of motion. This means that cosmological evolution of a flat anisotropic universe
in $R+R^2$ gravity needs a special investigation which is the matter of the present paper.

The paper is organized as follows: In section 2, we present the basic equations of the model. In section 3 we describe schematically the strategy adopted to obtain the correct degrees of freedom and then we analyze the dynamics of $R^2$-gravity both in the vacuum case and in the case with matter; we find exact solutions and determine their stability. In section 4, we derive the analytic behavior of the shear using a general line element. Finally, section 5 contains a summary of the results and conclusions.

\section{System under consideration}
\label{sec-sys}

The gravitational action considered in our analysis is the following
\begin{equation}
S=\frac{1}{16\pi G}\int d^{4}x \sqrt{-g}\left[\left(R-2\Lambda\right)+\beta R^{2}\right]\,,
\label{acao}
\end{equation}
where $g$ is the determinant of the metric, $G$ the Newton constant and $\beta$ a parameter.\\ This theory can be interpreted as a particular form of $f(R)$ gravity. 
Observations tell us that the dimensionless coefficient $\frac{\beta}{16\pi G}$ is very large, $\approx 5 \times 10^8$. This follows from the fact that its expression in terms of observable
quantities, in the leading order of the slow-roll approximation, is
$\frac{\beta}{16\pi G}= \frac{N^2}{288\,\pi^2 P_{\zeta}(k)}$ where $P_{\zeta}(k)$ is the power
spectrum of primordial scalar (adiabatic) perturbations, {\it N} is both 
$ \ln{k_f/k}$ and the number of e-folds from the end of inflation, $k_f$ being
the wave vector corresponding to the comoving scale equal to the Hubble 
radius at the end of inflation ($k_f/a(t_0)$ is slightly less than the CMB 
temperature now): see e.g.~\cite{Netto:2015cba}. For the $R+R^2$ inflationary model, $P_{\zeta}\propto N^2$, 
so $\beta$ is a constant indeed. Note also that $\beta=\frac{1}{6M^2}$ where $M$ is the scalaron mass after the end of inflaton (and in flat space-time, too). On the other hand, the coefficient of the 
Weyl square term (that is present in general quadratic model of gravity) in the Lagrangian density generated by one-loop quantum
gravitational effects is not expected to be so large. Typically it is of
the order of unity (or even significantly less due to small numerical
factors) multiplied by the number of elementary quantum fields. Thus,
there exists a large range of the Riemann and Ricci curvature where the
$R^2$ term dominates while the contribution from the
Weyl square term is still small. For this reason, anisotropic solutions
preceding the inflationary stage may be studied using the same $R+R^2$ model 
up to curvatures of the order of the Planck one.\\

Metric variation of the theory in (\ref{acao}) gives the following fields equations
\begin{equation}
E_{ab}\equiv\left(G_{ab}+g_{ab}\Lambda\right)+\beta H_{\: ab}^{(1)}=0\,,
\label{eq.campo}
\end{equation}
where
\begin{eqnarray}
 &  & G_{ab}=R_{ab}-\frac{1}{2}g_{ab}R\,,\nonumber\\
 &  & H_{ab}^{(1)}=-\frac{1}{2}g_{ab}R^{2}+2RR_{ab}+2g_{ab}\nabla^2 R-2R_{;ab}\,. \label{H1}
\end{eqnarray}
 $H_{ab}^{(1)}$ is the contribution coming from the variation of $R^2$ term. Let us emphasize that every Einstein metric satisfying $R_{ab}=g_{ab}\Lambda$
is an exact solution of (\ref{eq.campo}). This implies that all vacuum solutions of GR are also exact solutions of the quadratic theory \eqref{acao}. Any source that satisfies $\nabla^{c}T_{ca}=0$, can be consistently added to the right hand side of \eqref{eq.campo}.\\
As anticipated, a powerful tool to provide exact solutions of quadratic theory of gravity, is the dynamical system approach which allows for the determination of fixed point and for a qualitative description of the global dynamics of the system. It is particularly suited for the study of the dynamics of anisotropic spacetimes~\cite{wainwright2005dynamical}, like spatially homogeneous Bianchi metrics. In this case we can write the line-element as
\begin{eqnarray}
 &  & ds^{2}=-\frac{dt^{2}}{H(t)^{2}}+\delta_{ij}{\bf {\omega}}^{i}\otimes\omega^{j}\,,
 \end{eqnarray}
where the $i,\, j$ indices refer to the spatial part and $\omega^{j}$ is a triad of one-forms  
\begin{equation}
d \omega^{i}=-\frac{1}{2}C^{a}_{\,bc} \omega^{b}\otimes \omega^{c}\,,
\end{equation}
where $C^{a}_{\,bc}$ are the spatial structure constants of the Bianchi group, and depend only on time. They are usually defined as 
\begin{equation}
C^{a}_{\,bc}=\varepsilon_{bcd} n^{da}-\delta^{a}_{\,b} a_{c} + \delta^{a}_{\,c}a_{b}\,,
\end{equation} 
where the values of the symmetric matrix $n^{ab}$ and the vector $a_{b}$ define the various Bianchi models. In our case, where we focus on Bianchi I metric, we have $n_{ab}=0$ and $a_{b}=0$. \\
Defining the time-like vector $u^{a}=(H,0,0,0)$, which satisfies the normalization condition $u^{a}u_{a}=-1$, and the projection tensor $h_{ab}=g_{ab}+u_{a}u_{b}$, we can define the relevant kinematical quantities	\begin{eqnarray}
 &  & \nabla_{a}u_{b}=\sigma_{ab}+\omega_{ab}+\frac{1}{3}\theta\delta_{ab}-\dot{u}_{a}u_{b}\,,\nonumber \\
 &  & \sigma_{ab}=u_{(a;b)}-\frac{1}{3}\theta\delta_{ab}+\dot{u}_{(a}u_{b)}\,,\nonumber \\
 &  & \omega_{ab}=u_{[a;b]}+\dot{u}_{[a}u_{b]}\,,\nonumber \\
 &  & \dot{u}_{a}=u^{b}\nabla_{b}u_{a}\,,\nonumber \\
 &  & \theta=\nabla_{c}u^{c}\,,\label{teta}
\end{eqnarray}
where $\sigma_{ab}$ is the symmetric shear tensor $(\sigma_{ab}=\sigma_{(ab)}, \sigma_{ab}u^{b}=0, \sigma^{a}_{\;a}=0)$, $\omega_{ab}$ is the vorticity tensor $(\omega_{ab}=\omega_{(ab)}, \omega_{ab}u^{b}=0)$ and $\dot{u}_{a}$ is the acceleration vector. $\theta$ is the volume expansion, and it is related to the Hubble parameter by
\begin{equation}
\theta=\frac{1}{3}H\,.
\label{H}
\end{equation}
In our analysis we consider a cosmological model where the shear is diagonal and is defined as
\begin{eqnarray}
 &  & \sigma_{ij}=\mbox{diag}\left[-\frac{2\sigma_{+}}{H},\frac{\sigma_{+}+\sqrt{3}\sigma_{-}}{H},\frac{\sigma_{+}-\sqrt{3}\sigma_{-}}{H}\right],\label{sigma}
\end{eqnarray}
and since we will consider spatially homogeneous spacetimes,
the time-like vector is geodesic $\dot{u}^{a}=0$ with zero vorticity
$\omega_{ab}=0$, being normal to the time slices.\\

We consider a perfect fluid source with no anisotropic pressures  so the energy-momentum tensor is
\begin{equation}
T_{ab}=(\rho+p)u_au_b +pg_{ab}\,,
\end{equation}
and it can be decomposed schematically as 
\begin{equation}
8\pi GT_{ab}=\mbox{diag} [3\Omega_m,3wH^2\Omega_m,3wH^2\Omega_m,3wH^2\Omega_m]\,.
\label{emtensor}
\end{equation}
where $w$ is the equation of state (EoS) parameter.\\
In order to have a system of autonomous first order differential equations we divide the shear $\sigma_{\pm}$, given in \eqref{sigma}, and density parameters by appropriate powers of $H$, defining in this way the new dimensionless expansion-normalized variables (ENV) 
\begin{eqnarray}
 &  & \Sigma_{\pm}=\frac{\sigma_{\pm}}{H}\,,\nonumber \\
 &  & \Omega_{m}=\frac{8 \pi G \rho }{3H^{2}}\,,\nonumber \\
 &  & \Omega_{\Lambda}=\frac{\Lambda}{3H^{2}}.\label{Omegas}
\end{eqnarray}
where $\rho/H^2= T_{00}$ is the energy density.\\
The rest of the ENV are zero since we are restricting to the Bianchi I case. The time evolution of the sources follow directly from the conservation of the energy momentum tensor ($\nabla^{b} T_{ab}=0$) and from the definition itself of \eqref{Omegas},
\begin{eqnarray}
 &  & \dot{\Omega}_{m}=-3(w+1)\Omega_{m}-2Q_1\Omega_{m}\,,\nonumber\\
 &  & \dot{\Omega}_{\Lambda}=-2Q_1\Omega_{\Lambda}\,. 
 \label{e.t.Omegas}
\end{eqnarray}
The fact that we have higher order theory of gravity, requires the introduction of  
additional ENVs, which reflect the higher order time derivatives in the equations of motion, as firstly done in~\cite{Barrow:2006xb} 
\begin{eqnarray}
 &  & Q_{1}=\frac{\dot{H}}{H}\,,\nonumber\\
 &  & Q_{2}=\frac{\ddot{H}}{H^2}\,,\nonumber\\
 &  & B=\frac{1}{3\beta H^{2}}\,.
\label{Sigma}
\end{eqnarray}
According to their own definitions, these ENV must satisfy the following differential equations
\begin{eqnarray}
 &  & \dot{\Sigma}_{\pm}=\Sigma_{\pm1}-\Sigma_{\pm}Q_{1}\,,\nonumber\\
 &  & \dot{B}=-2Q_{1}B\,,\nonumber\\
 &  & \dot{Q}_{1}=Q_{2}-Q_{1}^{2}\,.\label{e.t.Sigma}
\end{eqnarray}
So now we have all the ingredients the compute the evolution of our theory. The complete dynamical system is  given by  the equations (\ref{e.t.Omegas}), (\ref{e.t.Sigma}) and by the differential equations shown in the Appendix~\ref{app-A}.

\section{Generalized anisotropic solutions}
\label{sec-gensol} 
\begin{figure}[t!]
    \centering 
   \includegraphics[scale=0.40, angle=0]{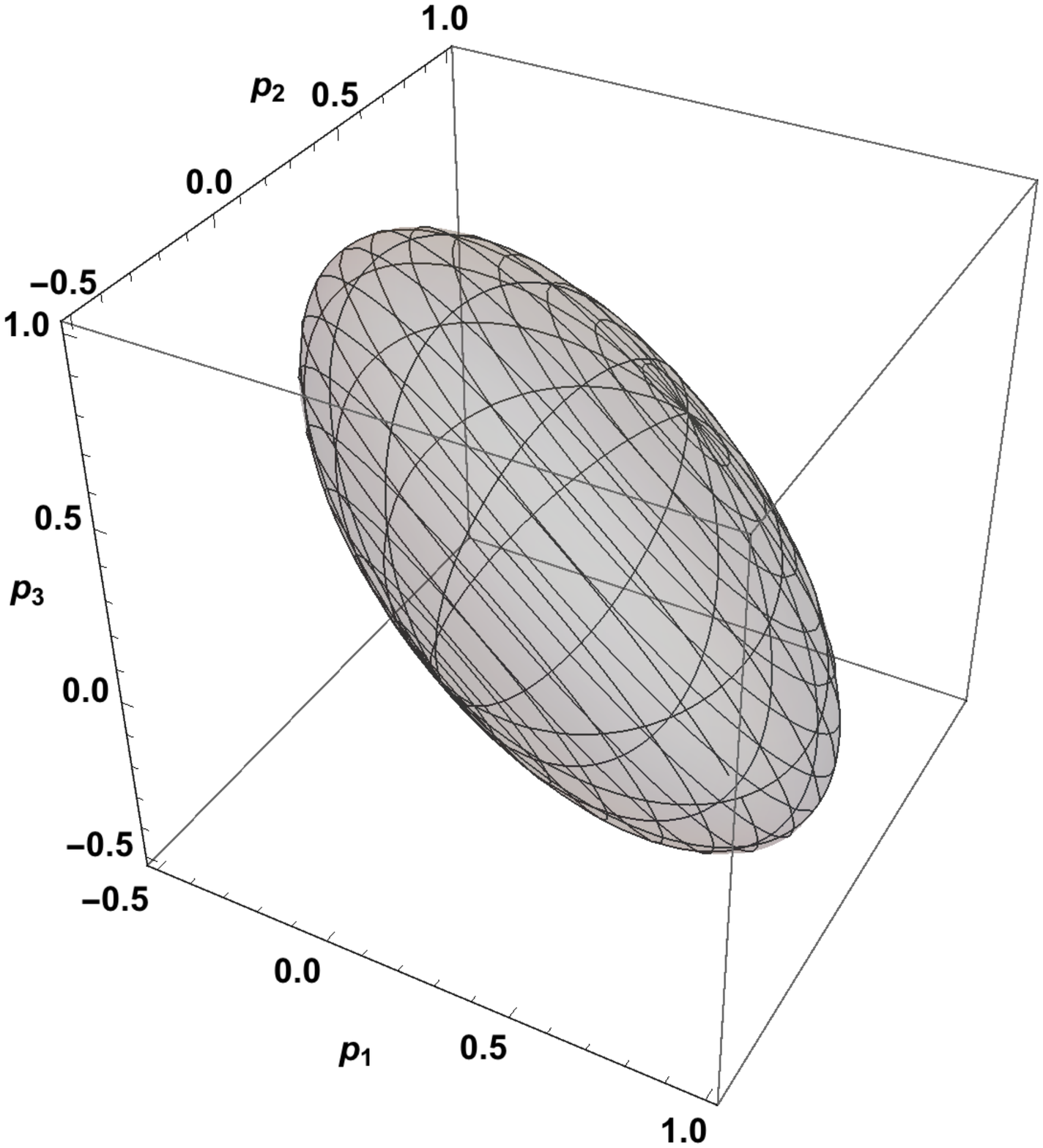}
    \caption{Ellipsoid in the parameter space $p_1$, $p_2$ and $p_3$ given by equation (\ref{ps}).}
    \label{ellipsoid}
\end{figure}

Now we start from this particular line element for the spacetimes
\begin{equation}
ds^{2}=-d\tau^{2}+\tau^{2p_{1}}dx^{2}+\tau^{2p_{2}}dy^{2}+\tau^{2p_{3}}dz^{2}.\label{le}
\end{equation}
For vanishing cosmological constant, and near the singularity when $\tau\rightarrow 0$, the Einstein tensor for the above line element goes like  $G_{ab}\sim 1/\tau^2$, so it becomes negligible in comparison to the $H^{(1)}_{ab}\sim 1/\tau^4$ given in (\ref{H1}). By direct substituting the line element (\ref{le})
into the field equations \eqref{eq.campo} for vacuum source, a purely algebraic
equation is obtained when $B=\frac{1}{3\beta H^2}\rightarrow 0$
\begin{equation}
p_{2}^{2}+p_{2}(-1+p_{1}+p_{3})-(p_{1}+p_{3}-p_{1}^{2}-p_{3}^{2}-p_{1}p_{3})=0\,.
\label{peq}  
 \end{equation}
 The set of solutions \eqref{peq} can be parametrized using two angles $\psi$ and $\phi$ 
\begin{eqnarray}
 &  & p_{1}=\sqrt{\frac{3}{8}}\sin\phi\left(\frac{\cos\psi+\sqrt{2}\sin\psi}{\sqrt{2}}\right)+\frac{1}{4}\,,\nonumber \\
 &  & p_{3}=\sqrt{\frac{3}{8}}\sin\phi\left(\frac{\cos\psi-\sqrt{2}\sin\psi}{\sqrt{2}}\right)+\frac{1}{4}\,,\nonumber \\
 &  & p_{2}=\frac{1}{4}-\sqrt{\frac{3}{8}}\sin\phi\frac{\cos\psi}{\sqrt{2}}+\sqrt{\frac{3}{8}}\cos\phi,\label{ps}
\end{eqnarray}
where $\psi=[0,2\pi]$ and $0<\phi<\pi, $ and lies in the surface of an ellipsoid shown in figure \ref{ellipsoid}. In this surface are contained both generalized Kasner solution and the isotropic solution. \\
    \\
The expansion normalized variables for the line element \eqref{le} read 
\begin{eqnarray}
 &  & Q_{1}=-\frac{3}{p_{1}+p_{2}+p_{3}}\,,\nonumber\\
 &  & Q_{2}=\frac{9}{(p_{1}+p_{2}+p_{3})^{2}}\,,\nonumber\\
 &  & \Sigma_{+}=\frac{-3p_{1}+(p_{1}+p_{2}+p_{3})}{2(p_{1}+p_{2}+p_{3})}\,,\nonumber\\
 &  & \Sigma_{-}=\frac{\sqrt{3}(p_{2}-p_{3})}{2(p_{1}+p_{2}+p_{3})}\,. 
\label{sol_env}
\end{eqnarray}
The solution space given by eq.~\eqref{peq} can be written in a more compact form using the variables $u=p_1^2+p_2^2+p_3^2$ and $s=p_1+p_2+p_3$,  with $\tau\rightarrow 0$
 \begin{equation}
 \frac{u}{2} +\frac{s^2}{2}-s=0\,,
 \label{eqcampovacuo}
 \end{equation}
 or with respect to ENV with $B\rightarrow 0$, as
 \begin{equation}
 2+Q_1+\Sigma_-^2+\Sigma_+^2=0.\label{sol.env}
 \end{equation}
 The solution of (\ref{eqcampovacuo}) is easily obtained as 
 \begin{equation}
u=2s-s^2\,.
\label{solucaovacuo}
\end{equation}
This compact way of writing the equation, is particular suitable to check the solutions: in fact it can be easily seen that Kasner ($s=1$ and $u=1$) and the isotropic vacuum ($s=3/2$ and $u=3/4$) are both particular solutions of this equation\footnote{The same generic behaviour near an anisotropic curvature singularity occurs for a non-minimally coupled scalar field in many cases,
in particular, for a massive conformally-coupled field, see the recent
paper~\cite{Kamenshchik:2017fk} in this connection.}. We also remember that, in terms of the ENV, generalized Kasner's solution is given by $Q_1=-3$, $\Sigma_+^2+\Sigma_-^2=1$, and the isotropic vacuum solution by $Q_1=-2$ and $\Sigma_+=\Sigma_-=0$. And that both of these solutions  belong to the solution set given by (\ref{eqcampovacuo}). 

The Ricci scalar can be written in terms of the ENV, and in terms of variables $s$ and $u$, like
\begin{eqnarray} 
&&R=\frac{2}{\beta B}\left(2+Q_{1}+\Sigma_{+}^{2}+\Sigma_{-}^{2}\right)\,,\nonumber\\
&&R=\frac{6}{\beta B}\left( \frac{s^2/2 +u/2-s}{s^2}\right)\,,
\label{Riemann_scalar}
\end{eqnarray}
such that the solution given in \eqref{solucaovacuo} ,\eqref{sol.env} as long as $B\ne0$, results in
\[
R=0\,. 
\]
By looking at~\eqref{H1}, it is not difficult to convince ourselves that zero Ricci scalar ($R=0$) is in fact the asymptotic solution of eq.~\eqref{eq.campo}. If near the singularity when $\tau\rightarrow 0$, the most important contribution to (\ref{eq.campo}) comes from $H^{(1)}_{ab}$ given in (\ref{H1}), the field equation can be approximated by 
\begin{equation}
H^{(1)}_{ab}\approx 0,
\end{equation}
then  $R=const.=0$ implies $H^{(1)}_{ab} = 0$.  Let us stress that this solution is only valid if the other terms in the field equation (\ref{eq.campo}) are negligible in comparison to $H^{(1)}_{ab}$. 

On the other hand for vanishing cosmological constant and absence of classical sources, in a non perturbative picture in the sense that the Einstein tensor $G_{ab}$ is not disregarded, it behaves as an effective source for field equations. And even though it diverges at the singularity (\ref{eqcampovacuo}), the following ratios of the effective pressures to energy densities obtained directly from (\ref{le})
\begin{eqnarray*}
&&\epsilon_1=\frac{G_1^1}{G_{00}}=-\frac{p_3^2-p_3+p_3p_2+p_2^2-p_2}{p_2p_1+p_3p_1+p_3p_2}\,,\\
&&\epsilon_2=\frac{G_2^2}{G_{00}}=-\frac{p_1^2-p_1+p_3^2-p_3+p_1p_3}{p_2p_1+p_3p_1+p_3p_2}\,,\\
&&\epsilon_3=\frac{G_3^3}{G_{00}}=-\frac{p_1^2-p_1+p_2^2-p_2+p_2p_1}{p_2p_1+p_3p_1+p_3p_2}\,,
\end{eqnarray*}
do have a well defined limit. The trace indicates that at the singularity, given by (\ref{eqcampovacuo}), the effective EOS parameter behaves as in radiation
\begin{equation}
\epsilon_1+\epsilon_2+\epsilon_3=1\,.
\end{equation}

\subsection{Stability analysis \label{stability}}
In the dynamical system approach, the field equations are re written with respect to the ENV, such that the solutions are fixed points.  In particular, the solution space described in the previous section constitute an invariant set of fixed points. The linearization around the fixed points reveals the local stability of the theory. In fact, since all eigenvalues $\lambda_i \geq 0$, this solution set is an attractor to the past, as all trajectories to the future deviate exponentially from this solution set. Stability with and without matter source is going to be addressed, and the presence of matter is irrelevant for sufficiently big shear. 
\subsubsection{Obtaining the dynamical system\label{dof}}
In order to describe the evolution of the correct degrees of freedom, in this section we will describe the strategy that we have adopted in order to simplify
the system of equations of motion. From the $E_{11}, E_{22}$ and $E_{33}$ equations in \eqref{eq.campo} we can isolate the variable related to the higher order time derivative $Q_{2}$; then we find a system of $3$ differential equations
\begin{eqnarray}
 &  & \dot{Q}_{2}=f_{1}(Q_{1},Q_{2},\Sigma_{\pm2},\Sigma_{\pm1},\Sigma_{\pm},\Omega,B)\,,\nonumber\\
 &  & \dot{Q}_{2}=f_{2}(Q_{1},Q_{2},\Sigma_{\pm2},\Sigma_{\pm1},\Sigma_{\pm},B)\,,\nonumber\\
 &  & \dot{Q}_{2}=f_{3}(Q_{1},Q_{2},\Sigma_{\pm2},\Sigma_{\pm1},\Sigma_{\pm},B)\,,
 \label{fieldeq}
\end{eqnarray}
where $f$ is a generic function of all the remaining ENV. Form the $E_{00}$ component of \eqref{eq.campo}, we obtain a constraint equation 
\begin{equation}
0=C_{1}(\Sigma_{\pm1},\Sigma_{\pm},Q_{1},Q_{2},\Omega,B)\,,
\label{constr}
\end{equation}
that, as we will see, will be important to check the stability of the numerical evolution of the dynamical system.\\
By doing a linear combinations of the field equations~\eqref{fieldeq}, we can obtain two additional
constraints that read as 
\begin{eqnarray}
 &  & 0=C_{2}(\Sigma_{\pm1},\Sigma_{\pm},Q_{1},Q_{2},\Omega,B)\,,\nonumber\\
 &  & 0=C_{3}(\Sigma_{\pm1},\Sigma_{\pm}Q_{1},Q_{2},\Omega,B)\,.
 \label{constr2}
\end{eqnarray}
So now, from the constraints~\eqref{constr} and~\eqref{constr2}, it is possible to write three algebraic equation for $Q_{2}$, $\Sigma_{+1}$ and $\Sigma_{-1}$, that will be function of the remaining variables and we write schematically as
\begin{eqnarray}
 &  & Q_{2}(\Sigma_{\pm},Q_{1},\Omega,B)\,,\nonumber \\
 &  & \Sigma_{+1}(\Sigma_{\pm},Q_{1},\Omega,B)\,,\nonumber \\
 &  & \Sigma_{-1}(\Sigma_{\pm},Q_{1},\Omega,B)\,.
\label{Q2S1}
\end{eqnarray}
If now, we consider the ENV related to $\ddot{\sigma}_{\pm}$, defined as $\Sigma_{\pm 2}=\frac{\ddot{\sigma}_{\pm}}{H}$, we can use its definition to derive the equation 
 $\dot{\Sigma}_{\pm 1}=\Sigma_{\pm2}-Q_{1}\Sigma_{\pm1}$. Using~\eqref{Q2S1}, it is now possible
 to derive the equation for $\Sigma_{\pm2}$
 \begin{equation}
\Sigma_{\pm2}(\Sigma_{\pm},Q_{1},\Omega,B)\,.
\end{equation}
Substituting the last equation and eq.~\eqref{Q2S1} into
the original dynamical system equations we finally obtain also the equation for  
$\dot{Q}_{2}=f_{1}(\Sigma_{\pm},Q_{1},\Omega,B)$.\\

Then  the complete dynamical system is described by the following equations 
\begin{eqnarray}
 &  & \dot{Q}_{1}=Q_{2}(\Sigma_{\pm},Q_{1},\Omega,B)-Q_{1}^{2}\,,\nonumber \\
 &  & \dot{\Sigma}_{\pm}=\Sigma_{\pm1}(\Sigma_{\pm},Q_{1},\Omega,B)-Q_{1}\Sigma_{\pm}\,,\nonumber \\
 &  & \dot{B}=-2Q_{1}B\,,\nonumber \\
 &  & \dot{\Omega}_{m}=\left(-2Q_{1}-3(w+1)\right)\Omega_{m}\,,\nonumber \\
 &  & \dot{Q}_{2}=f_{1}(\Sigma_{\pm},Q_{1},\Omega,B)\,.
 \label{dynsyst}
\end{eqnarray}
where in our analysis the last equation will be integrated numerically to be compared with the algebraic
relation $Q_{2}(\Sigma_{\pm},Q_{1},\Omega,B)$ contained in~\eqref{Q2S1}. Moreover
we will use one of the constraints to obtain a conserved quantity 
to numerically check the stability of our results. The last equation is not a dynamical degree
of freedom, but just an artifact to check numerically the solutions. \\ 

Looking at the above set of equations, it can be noted that there is only
one additional dynamical degree of freedom compared to General Relativity, which
is the first equation for $\dot{Q}_{1}$. This can be easily understood by remembering that, through 
a conformal transformation, this gravitational theory is equivalent
to GR plus a scalar field~\cite{Sotiriou:2008rp}.

 As we have described above, the linearization of the dynamical system (\ref{dynsyst}) around the solution (\ref{sol.env}) gives rise to the following eigenvalues, that will be discussed in the next subsections.
\subsubsection{Pure geometric modes}

\begin{figure}[t!]
    \centering 
    \captionsetup{width=.8\linewidth}
    \begin{tabular}{c c}
    \resizebox{0.44\columnwidth}{!}{\includegraphics{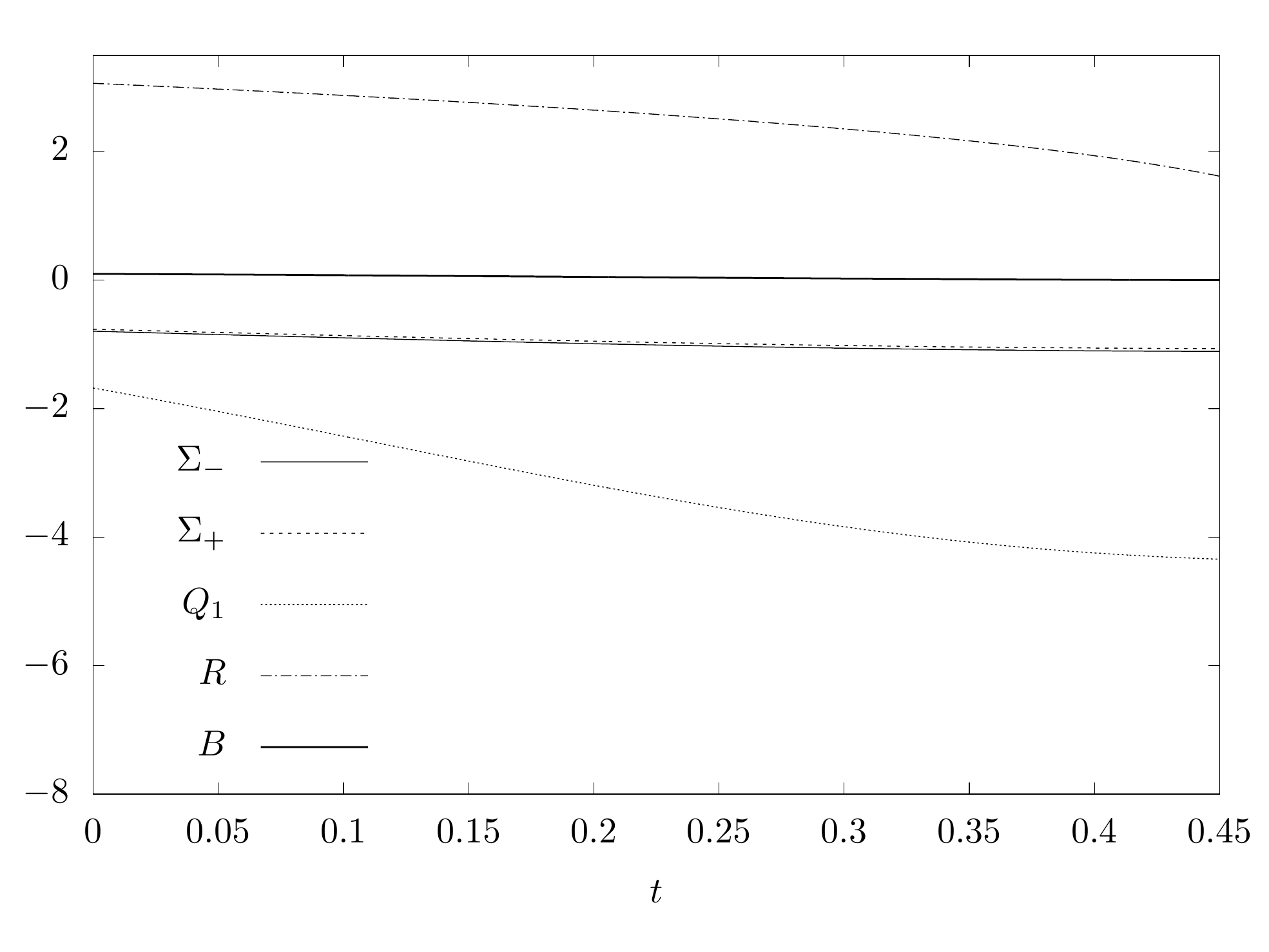}} & 
    \resizebox{0.45\columnwidth}{!}{\includegraphics{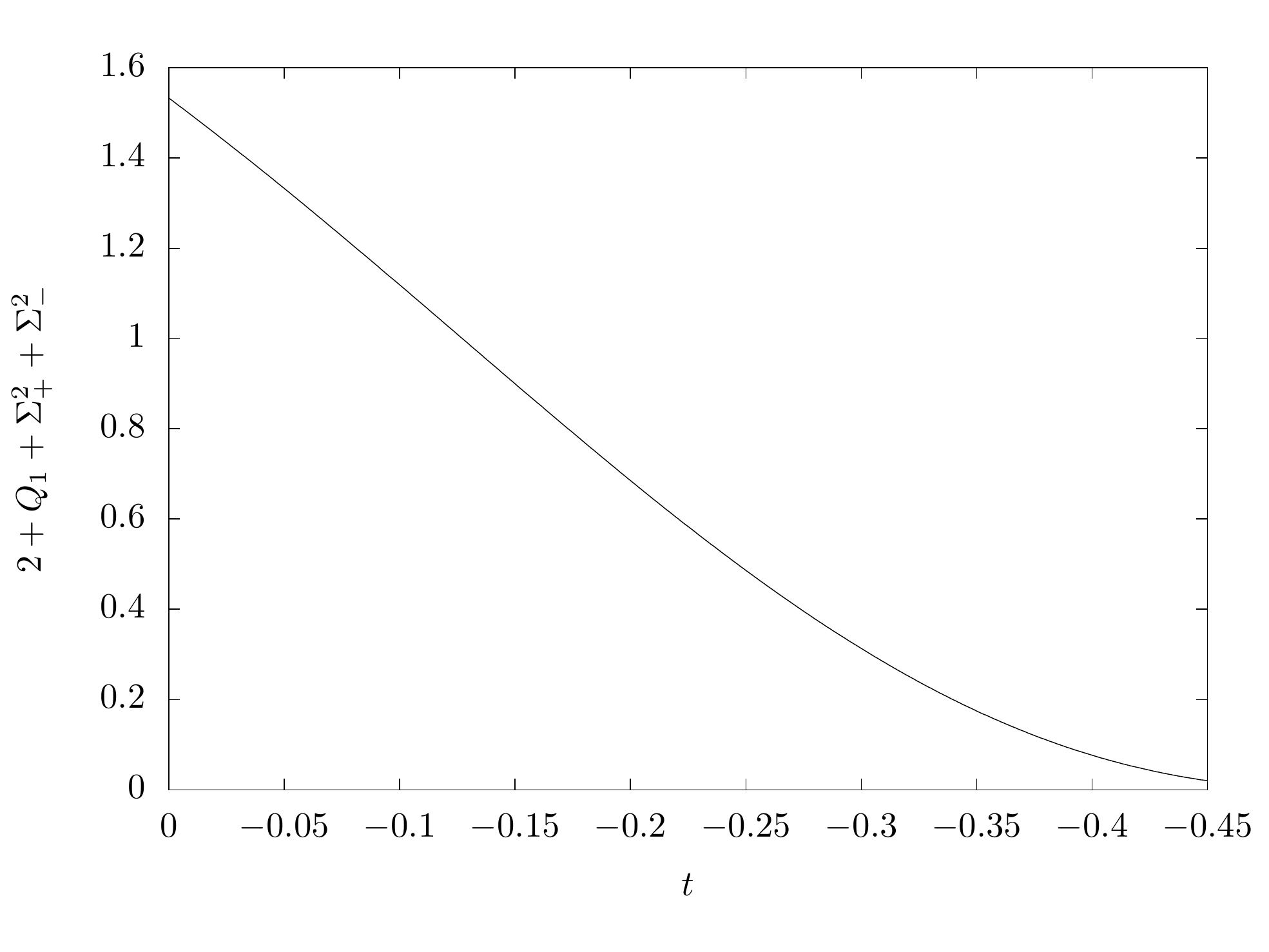}}\\
    $a)$ & $b)$ 
    \end{tabular}
     \caption{Backwards time evolution, showing the approach to the singularity. The expansion normalized shear  variables $\Sigma_+$ and $\Sigma_-$ and $Q_1$ approach constant asymptotic values. We explicitly checked that asymptotically $Q_1<-4.37$. In $a)$ it can be seen that the Ricci scalar decreases which is consistent with eq. \eqref{ev_ricci_scalar} when $Q_1<-3$. In $b)$ it is explicitly shown that the numerical solution approaches eq. \eqref{sol_env}, as it should for all points of the past attractor. Even if not reported in the plot, the constraint has been checked to be satisfied numerically.}
     \label{backevolution}
    \end{figure}

As a first case we consider the vacuum case (without the matter modes). In this case the stability of the system is characterized by the following eigenvalues 
\begin{eqnarray}
& &\lambda_{1}=2(2+\Sigma_{-}^2+\Sigma_{+}^2)\,,\nonumber\\
& &\lambda_{2}=3(\Sigma_{+})^{2}+3(\Sigma_{-})^{2}+3\,,\nonumber\\
& & \lambda_{3}=0_{2}\,.
\label{vacuumeigen}
\end{eqnarray}
Except for the two zero eigenvalues this solution is an attractor to the past. These two zero 
eigenvalues appears naturally because in fact we have two-dimensional set of fixed points So, according to this theory the universe began as a generalized non-isotropic solution as shown in Figure \ref{backevolution}. There it can be seen in panel $b)$  that an arbitrary initial condition approaches eq. \eqref{sol_env} which defines this solution set. Even if non reported the constraint was numerically verified.


\subsubsection{Matter modes}

Allowing perturbations in the matter sector $(\Omega_{\Lambda},\,\Omega_{m})$
we have the following eigenvalues, 
\begin{eqnarray}
 &  & \lambda_{1}=3(\Sigma_{+})^{2}+3(\Sigma_{-})^{2}+3\,,\nonumber\\
 & & \lambda_{2}=1-3w+2(\Sigma_{+})^{2}+2(\Sigma_{-})^{2}\,,\nonumber\\
 & & \lambda_{3}=0_{2}\,,\nonumber\\
 &  & \lambda_{4}=(4+2(\Sigma_{+})^{2}+2(\Sigma_{-})^{2})_2\,.
\end{eqnarray}
Excluding the zero eigenvalues, it is interesting that this solution
is an attractor to the past in the presence of cosmic substance for
any initial values for $\Sigma_{+}$ and $\Sigma_{-}$ as long as
the EoS parameter $-1<w<1/3$. For $1/3<w<1$, the solution will still
be an attractor to the past for sufficiently big initial values of
$\Sigma_{+}$ and $\Sigma_{-}$ such that $\lambda_{2}$ is positive. 

We focused on solutions which are attractors to the past, however there can be other solutions~\cite{Barrow:2006xb}.


\section{Analytic behavior}
\label{sec-analbe}
Now, starting from a remarkable property of any $f(R)$ gravity, obtained in~\cite{Gurovich:1979xg}, it is possible to determine  the dynamical evolution of the shear as an exact, analytical result. Considering a Lagrangian like
\begin{equation}
\mathcal{L}=\sqrt{-g}f(R)\,,
\label{effeR} 
\end{equation}
the variation with respect to the metric gives the following field equation 
\begin{equation}
f^\prime R_{ab} -\frac{1}{2} f g_{ab} -\nabla_a\nabla_b f^\prime + \nabla ^2 f^\prime g_{ab}=\kappa T_{ab}\,,
\label{fr}
\end{equation}
where $f'$ is used to denote derivative wrt to $R$, and $'$ represents derivative wrt to the proper time. $T_{ab}$ is the energy momentum tensor of some matter source. Since we are considering a spatially homogeneous spacetime, with corresponding sources, the following combinations are zero $a)\,T_{22}-T_{33}=0$, $b)\,T_{11}-T_{22}/2-T_{33}/2=0$. In the same way, taking the same combinations of the left hand side of the field equations (\ref{fr}), we have
\begin{eqnarray} 
&&a)\;2H\sqrt{3}\left(\frac{d}{dt}\sigma_- +3\sigma_-\right)f^\prime+2H\sqrt{3}\sigma_-\frac{d}{dt}f^\prime=0\,,\\
&&b)\; 3H\left(\frac{d}{dt}\sigma_+ +3\sigma_+\right)f^\prime+3 H\sigma_+\frac{d}{dt}f^\prime=0\,.
\end{eqnarray}
We can see that these equations admit as solutions 
\begin{eqnarray} 
&&a)\;\sigma_-=\frac{C_- e^{-3t}}{f^\prime}\,,\\
&&b)\;\sigma_+=\frac{C_+ e^{-3t}}{f^\prime}\,. 
\label{sigmafr}
\end{eqnarray}
Now we can connect the constants ($C_{\pm}$) with the parameters $p_1, p_2$ and $p_3$. As in~\cite{Gurovich:1979xg} we can write the scale factors given in (\ref{le}) as 
\begin{equation} 
a=rg_1,\;\; b=rg_2,\;\;c=rg_3,\;\;\;\;\; \mbox{with} \;\;\;\;\; abc=r^3,
\end{equation}
such that assuming $\prod_ig_i=1$, we have
\begin{equation}
\frac{\dot{g}_1}{g_1}+\frac{\dot{g}_2}{g_2}+\frac{\dot{g}_3}{g_3}=0.
\end{equation}
Taking into account~\eqref{teta}, \eqref{H} and \eqref{sigma} we find that
\begin{eqnarray}
&&\frac{\dot{r}}{r}=H\,,\\
&&\frac{\dot{g}_1}{g_1}=-2\sigma_+,\;\;\;\frac{\dot{g}_2}{g_2}=\sigma_+ +\sqrt{3}\sigma_-,\;\;\; \frac{\dot{g}_3}{g_3}=\sigma_+- \sqrt{3}\sigma_-\,.
\label{gs}
\end{eqnarray}
Remembering that derivatives with respect to proper time are related to derivatives with respect to dynamical time by $\frac{d}{d\tau}=\frac{dt}{d\tau} \frac{d}{dt}=H\frac{d}{dt}$, we can solve the first of~\eqref{gs} 
\begin{equation}
r=e^{t}\,,
\end{equation}
which, when substituted into \eqref{sigmafr} and \eqref{gs}, gives
\begin{equation}
\frac{\dot{g}_i}{g_i}=\frac{C_i}{r^3f^\prime}\,.
\end{equation}
This relation found here is the same of $(10)$ in~\cite{Gurovich:1979xg} with the constant $C_{\pm}$ related to $C_1$, $C_2$ and $C_3$ by
\begin{equation}
 C_1=-2C_+,\;\; C_2=C_+ +\sqrt{3}C_-,\; \;C_3=C_+-\sqrt{3}C_-\,.
\end{equation}
\\
When we consider the asymptotic solution discussed above we finally find that the relation with the coefficient $p_1$, $p_2$ and $p_3$ is the following
\begin{equation}
C_1=p_1-\frac{s}{3}, \;\;C_2=p_2-\frac{s}{3}, \;\; \mbox{and } C_3=p_3-\frac{s}{3}\,,
\end{equation}
where again $s=p_1+p_2+p_3$ and (\ref{solucaovacuo}) must be satisfied. $C_\pm$ can also be expressed as 
\begin{eqnarray}
&&C_+=\frac{-2p_1+p_2+p_3}{6}\,,\nonumber\\
&&C_-=\frac{3p_2-p_3}{6\sqrt{3}}\,.\label{c-c+}
\end{eqnarray}
It is now possible to understand the space of solutions close to the singularity, when $B\rightarrow 0$, as long as the solution stays near the asymptotic solution given in (\ref{eqcampovacuo}), which must be fulfilled since as explained in section \ref{stability} the solution is an attractor to the past. 
 
First of all, by taking the trace of (\ref{eq.campo}) bearing in mind (\ref{H1}), gives the following equation for $R$ in absence of sources 
\begin{eqnarray}
&&-R+6\beta\Box R=0\,,\nonumber\\
&&-R-\frac{1}{3\beta B}(\ddot{R}+(Q_1+3)\dot{R})=0 \label{ev_R}\,. 
\end{eqnarray}
Near the singularity, when $t \rightarrow -\infty$, $Q_1=-3/s=const.$ and $B=\frac{e^{6t/s}}{3 \beta H_0^2 }$ and eq. (\ref{ev_R}) 
\begin{equation}
\ddot{R}+(-3/s+3)\dot{R}+R\frac{e^{6t/s}}{H_0^2 }=0\,,
\label{Ricciev}
\end{equation}
has an analytical solution given by Bessel function of the first type $J_a$ and the Bessel function of the second type $Y_a$, which when written with respect to proper time $\tau$, $\exp(3t/s)\propto\tau$ is given by    
\begin{equation}
R=\{\hat{C}_1J_{(s-1)/2}(s\tau/(3H_0))+\hat{C}_2Y_{(s-1)/2}(s\tau/(3H_0))\}\tau^{(1-s)/2}.
\end{equation}
Asymptotically, as $\tau\rightarrow 0$ and $B\rightarrow 0$, \eqref{Ricciev} simplifies as
\begin{equation}
\ddot{R}+(-3/s+3)\dot{R}=0\,,
\end{equation}
and when $s\ne 1$ the solution is
\begin{equation}
R=C
+C_0\exp [-(Q_1+3)t]=\frac{C}{3(1-s)}+\tilde{C} \tau^{(1-s)}\,,
\end{equation}
while for $s=1$, which means $Q_1=-3$, it is
\begin{equation}
R=C_1+\tilde{C_1}t=C_2+\tilde{C_2}\frac{s}{3}\ln(\tau), 
\end{equation}
where all the $C$s are constants. 

This asymptotic behavior of $R$ can be substituted into \eqref{sigmafr} for the particular theory analyzed hitherto,  for which we have $f^\prime=1\,+\,2\beta R$, 
\begin{eqnarray}
&&\Sigma_+=\frac{\sigma_+}{H}=\left( \frac{1}{\exp(Q_1 t)}\right)\frac{C_+ e^{-3t}}{1+2\beta(C+C_0\exp [-(Q_1+3)t] )}	\,,\nonumber\\
&&\Sigma_-=\frac{\sigma_-}{H}=\left( \frac{1}{\exp(Q_1 t)}\right)\frac{C_- e^{-3t}}{1+2\beta(C+C_0\exp [-(Q_1+3)t] )}\,. \label{sigmaCs}
\end{eqnarray}
When $Q_1>-3$, which corresponds to $s>1$, this expression gives at the singularity ($t\rightarrow -\infty $)
\begin{eqnarray}
&&\Sigma_+=\frac{C_+}{2\beta C_0}\,,\\
&&\Sigma_-=\frac{C_-}{2\beta C_0}\,,
\end{eqnarray}
and since $2+Q_1+\Sigma_-^2+\Sigma_+^2=0$ with $Q_1=-3/s=const.$, it results in the following asymptotic form for the Ricci scalar
\begin{equation}
R=C+\frac{\sqrt{C_-^2+C_+^2}}{2\beta\sqrt{-Q_1-2}}\exp [-(Q_1+3)t]\,,
\end{equation}
and since the Ricci scalar must be real then $Q_1<-2$ which gives $s<3/2$.

We can also obtain the constant $C$ since we know that, interchangeably when $s<1$ or $Q_1<-3$, the asymptotic solution set \eqref{solucaovacuo} must continue to be a past attractor, see section \ref{stability}. This attractor has constant well defined values for $Q_1$, $\Sigma_\pm$ satisfying \eqref{sol_env} and when $Q_1<-3$ this will only occur if there is a particular cancellation in the denominator of \eqref{sigmaCs} $f^\prime\rightarrow 0$ giving a well defined limit for $\Sigma_\pm$ at the singularity 
\begin{equation}
1+2\beta C =0 \rightarrow C=-\frac{1}{2\beta}\,.
\end{equation}
We have the final asymptotic form for the Ricci scalar 
\begin{equation}
R=-\frac{1}{2\beta}+\frac{\sqrt{C_-^2+C_+^2}}{2\beta\sqrt{-Q_1-2}}\exp [-(Q_1+3)t]\,, 
\end{equation}
which is valid through $0<s<3/2$. Through (\ref{c-c+}), this last expression can be written as 
\begin{equation}
R=-\frac{1}{2\beta}+\frac{\sqrt{9p_1^2-9p_1p_2-9p_1p_3+9p_2^2+3p_3^2}}{18\beta\sqrt{3/s-2}}\exp [-(-3/s+3)t]\,.\label{ev_ricci_scalar}
\end{equation}
The numerical behavior of the Ricci scalar is shown in Figure \ref{backevolution} in panel $a)$. There it can be seen that the asymptotic constant value $R \simeq-1/(2\beta)$ is not reproduced exactly in the plot. The reason for that is due to the fact that,  in the dynamical system described in the appendix, the following denominator $4\Sigma_{+}^{2}+4\Sigma_{-}^{2}+4Q_{1}+B+8$ occurs in all equations and it vanishes on the attractor set. Although it is not possible to check numerically the asymptotic time evolution of the Ricci scalar it is possible to see in Figure (\ref{backevolution}) panel $a)$ that the Ricci scalar decreases asymptotically as it is expected for $Q_1<-3$. 
\section{Conclusions} 
\label{sec-concl}

In the present paper we have considered the past attractor solution for the evolution
of the flat anisotropic universe in $R+R^2$ gravity. Our results, in combination with 
already known results, indicate that the properties of the universe evolution near a 
cosmological singularity change significantly taking into account anisotropy
and/or modifications of gravity. Indeed, the evolution of isotropic universe is determined
solely by the matter equation of state. When anisotropy is taken into account, 
this isotropic solution becomes future asymptotic solution, while generalized vacuum Kasner solution
becomes a past attractor (except for stiff fluid with Jacobs solution). 

In general quadratic gravity without anisotropy new vacuum isotropic solution (``false radiation'' solution)
is stable to the past. The anisotropic case has instead has two sub-cases, because general quadratic
corrections to the gravitational action has two independent terms which can be chosen as proportional to squares of scalar curvature and the Weyl tensor. In a general situation, when
these two terms are of the same order, the dynamical system describing the universe past 
evolution has both ``false radiation" and generalized Kasner solution as attractors (the latter is,
more precisely, a saddle-node fixed point). So the nature of cosmological singularity (isotropic or anisotropic) depends on initial conditions imposed.

However, since the $R+R^2$ inflationary model is observationally well
motivated, and we have argued in Sec. II above that there exists a large
range of the Riemann and Ricci curvatures where the anomalously large
$R^2$ term dominates the Einstein term $R$, while
a ``normal-size" Weyl squared term is still small, one can expect that new
solution appears. It has two parameters (so it is a two-dimensional set of solutions), and 
includes both isotropic ``false radiation'' and generalized anisotropic Kasner solution (which is a one-dimensional set) as subsets. Moreover, in some sense, it interpolates between them, because it is possible to construct line of solutions with one end being isotropic solution and the other end being a point in the generalized Kasner set.

All this intermediate points disappears when the correction proportional to Weyl square is added to the action,
leaving only isotropic and generalized Kasner solutions and this represents one of the main results of our paper. However, since $R^2$ inflation model is observationally well motivated we can neglect the coefficient in front of the Weyl term, and  we can expect that the two-dimensional set of solutions discussed in this paper could be a good approximation for realistic models in quadratic gravity.

In the present paper we have restricted the analysis to flat metrics. However positive spatial curvature could, 
in principle, destroy this regime and generate more complicated behavior similar to the Belinsky-Khalatnikov-Lifshitz (BKL)~\cite{Belinsky:1970ew} singularity in General Relativity. We leave this problem for future analysis.

\subsection*{Acknowledgements}
We are delighted to thank Sigbj\o rn Hervik for illuminating discussions and comments. D. M. and A. T. thank the University of Stavanger for the warm hospitality when this paper was started. D. M\"uller would like to thank the Brazilian agency FAPDF process no. 193.000.181/2016 for partial support. 
The work of A. S. and A. T. was supported by the RSF grant 16-12-10401.
The computations performed in this paper have been partially done with Maple 16 and with the \textit{Ricci.m} package for Mathematica.
For numerical codes we used GNU/GSL ode package, explicit embedded Runge-Kutta Prince-Dormand on Linux.

\begin{appendix}
\section{Appendix A}
\label{app-A}
 Dynamical system   equations in the case without matter  ($\Omega_{m}=\Omega_{\lambda}=0$):

\begin{eqnarray}
 &  & \dot{Q}_{1}=-Q_{1}^{2}-\left\{ -288\Sigma_{-}^{2}-288\Sigma_{-}^{4}-72\Sigma_{-}^{6}+8B+B^{2}-
 12Q_{1}\Sigma_{+}^{2}B-12Q_{1}\Sigma_{-}^{2}B\right.\nonumber\\
 &  &-240\Sigma_{-}^{2}Q_{1}\Sigma_{+}^{2}-36\Sigma_{-}^{2}\Sigma_{+}^{2}B
  -216\Sigma_{-}^{2}\Sigma_{+}^{4}-576\Sigma_{-}^{2}\Sigma_{+}^{2}-32\Sigma_{-}^{2}B-192Q_{1}\Sigma_{-}^{2}\nonumber\\
 & & -40Q_{1}^{2}\Sigma_{-}^{2}-18\Sigma_{-}^{4}B-120\Sigma_{-}^{4}Q_{1}-216\Sigma_{-}^{4}\Sigma_{+}^{2}
  -40Q_{1}^{2}\Sigma_{+}^{2}-32B\Sigma_{+}^{2}\nonumber\\
  & &-18B\Sigma_{+}^{4}-120Q_{1}\Sigma_{+}^{4}
 -192Q_{1}\Sigma_{+}^{2}+2Q_{1}^{2}B+16Q_{1}B+64Q_{1}^{2}+96Q_{1}+8Q_{1}^{3}\nonumber\\
 &  & -\Sigma_{+}^{2}B^{2}-288\Sigma_{+}^{4}-288\Sigma_{+}^{2}-72\Sigma_{+}^{6}
 \left.-\Sigma_{-}^{2}B^{2}\right\} /\left\{ 4\left(4\Sigma_{+}^{2}+4\Sigma_{-}^{2}+4Q_{1}+B+8\right)\right\}\,,\\
 \nonumber\\
 &  & \dot{\Sigma}_{+}=-\Sigma_{+}Q_{1}-\Sigma_{+}\left\{ 24\Sigma_{+}^{2}+24+16Q_{1}+2Q_{1}^{2}+24\Sigma_{-}^{2}
  +B\Sigma_{+}^{2}+6\Sigma_{-}^{4}+\Sigma_{-}^{2}B\right. \nonumber\\
 & &+12\Sigma_{-}^{2}\Sigma_{+}^{2}+8Q_{1}\Sigma_{+}^{2} \left.+2B+6\Sigma_{+}^{4}+8Q_{1}\Sigma_{-}^{2}\right\} /\left\{ 4\Sigma_{+}^{2}+4\Sigma_{-}^{2}+4Q_{1}+B+8\right\}\,, \\
 \nonumber\\
 &  & \dot{\Sigma}_{-}=-\Sigma_{-}Q_{1}-\Sigma_{-}\left\{ 24\Sigma_{+}^{2}+24+16Q_{1}+2Q_{1}^{2}+24\Sigma_{-}^{2}
   +2B+6\Sigma_{+}^{4}+B\Sigma_{+}^{2}\right.\nonumber\\
   & &+6\Sigma_{-}^{4}+\Sigma_{-}^{2}B+12\Sigma_{-}^{2}\Sigma_{+}^{2} \left.+8Q_{1}\Sigma_{+}^{2}+8Q_{1}\Sigma_{-}^{2}\right\} /\left\{ 4\Sigma_{+}^{2}+4\Sigma_{-}^{2}+4Q_{1}+B+8\right\} \,.
\end{eqnarray}

 Dynamical system   equations in the case with matter  ($\Omega_{m}\neq0$ and $\Omega_{\lambda}\ne0$):
\begin{eqnarray}
 &  & \dot{Q}_{1}=-Q_{1}^{2}-\left\{ -\Omega_{\lambda}B^{2}-4Q_{1}B\Omega_{m}-12\Sigma_{+}^{2}B\Omega_{m}-12\Sigma_{-}^{2}B\Omega_{m}
  -12Q_{1}\Sigma_{+}^{2}B-12Q_{1}\Sigma_{-}^{2}B\right.\nonumber\\
 & &-240\Sigma_{-}^{2}Q_{1}\Sigma_{+}^{2}-
 36\Sigma_{-}^{2}\Sigma_{+}^{2}B-4Q_{1}\Omega_{\lambda}B-12\Sigma_{+}^{2}\Omega_{\lambda}B-12\Sigma_{-}^{2}\Omega_{\lambda}B
 -B^{2}\Omega_{m}-216\Sigma_{-}^{2}\Sigma_{+}^{4}\nonumber\\
 & &-576\Sigma_{-}^{2}\Sigma_{+}^{2}-32\Sigma_{-}^{2}B-192Q_{1}\Sigma_{-}^{2}-40Q_{1}^{2}\Sigma_{-}^{2}-18\Sigma_{-}^{4}B
 -120\Sigma_{-}^{4}Q_{1}-216\Sigma_{-}^{4}\Sigma_{+}^{2}\nonumber\\
 & &-40Q_{1}^{2}\Sigma_{+}^{2}
-32B\Sigma_{+}^{2}-18B\Sigma_{+}^{4}-120Q_{1}\Sigma_{+}^{4}-192Q_{1}\Sigma_{+}^{2}
 +2Q_{1}^{2}B+16Q_{1}B-8\Omega_{\lambda}B\nonumber\\
 & &-8B\Omega_{m}+64Q_{1}^{2}+96Q_{1} +8Q_{1}^{3}-\Sigma_{+}^{2}B^{2}-288\Sigma_{+}^{4}-288\Sigma_{+}^{2}-72\Sigma_{+}^{6}
 -\Sigma_{-}^{2}B^{2}\nonumber\\
 & &-288\Sigma_{-}^{2}-288\Sigma_{-}^{4}-72\Sigma_{-}^{6} \left.+8B+B^{2}\right\} /\left\{ 4\left(4\Sigma_{+}^{2}+4\Sigma_{-}^{2}+4Q_{1}+B+8\right)\right\}\,, \\
 \nonumber\\
 &  & \dot{\Sigma}_{+}=-\Sigma_{+}Q_{1}-\Sigma_{+}\left\{ 24\Sigma_{+}^{2}+24+24\Sigma_{-}^{2}+2B+16Q_{1}+2Q_{1}^{2}
  +6\Sigma_{+}^{4}+B\Sigma_{+}^{2}+6\Sigma_{-}^{4}\right.\nonumber\\
 & &+\Omega_{\lambda}B+\Sigma_{-}^{2}B+12\Sigma_{-}^{2}\Sigma_{+}^{2}
  \left.+8Q_{1}\Sigma_{+}^{2}+B\Omega_{m}+8Q_{1}\Sigma_{-}^{2}\right\} /\left\{ 4\Sigma_{+}^{2}+4\Sigma_{-}^{2}+4Q_{1}+B+8\right\}\,, \\
\nonumber \\
 &  & \dot{\Sigma}_{-}=-\Sigma_{-}Q_{1}-\Sigma_{-}\left\{ 24\Sigma_{+}^{2}+24+24\Sigma_{-}^{2}+2B+16Q_{1}+2Q_{1}^{2}+6\Sigma_{+}^{4}
 +B\Sigma_{+}^{2}+6\Sigma_{-}^{4}\right.\nonumber\\
 & &+\Omega_{\lambda}B+\Sigma_{-}^{2}B+12\Sigma_{-}^{2}\Sigma_{+}^{2}+8Q_{1}\Sigma_{+}^{2} 
 \left.+B\Omega_{m}+8Q_{1}\Sigma_{-}^{2}\right\} \left\{ 4\Sigma_{+}^{2}+4\Sigma_{-}^{2}+4Q_{1}+B+8\right\} \,.
\end{eqnarray}

\end{appendix}

\newpage

\bibliographystyle{utcaps}
\bibliography{refsR2.bib}

\providecommand{\href}[2]{#2}\begingroup\raggedright\begin{thebibliography}{10}

\bibitem{Starobinsky:1987zz}
A.~A. Starobinsky and H.~J. Schmidt, ``{On a general vacuum solution of
  fourth-order gravity},''
\href{http://dx.doi.org/10.1088/0264-9381/4/3/026}{{\em Class. Quant. Grav.}
  {\bfseries 4} (1987) 695--702}.

\bibitem{Ade:2015lrj}
{\bfseries Planck} Collaboration, P.~A.~R. Ade {\em et~al.}, ``{Planck 2015
  results. XX. Constraints on inflation},''
  \href{http://dx.doi.org/10.1051/0004-6361/201525898}{{\em Astron. Astrophys.}
  {\bfseries 594} (2016) A20},
\href{http://arxiv.org/abs/1502.02114}{{\ttfamily arXiv:1502.02114
  [astro-ph.CO]}}.

\bibitem{Starobinsky:1980te}
A.~A. Starobinsky, ``{A New Type of Isotropic Cosmological Models Without
  Singularity},''
\href{http://dx.doi.org/10.1016/0370-2693(80)90670-X}{{\em Phys. Lett.}
  {\bfseries 91B} (1980) 99--102}.

\bibitem{Mukhanov:1990me}
V.~F. Mukhanov, H.~A. Feldman, and R.~H. Brandenberger, ``{Theory of
  cosmological perturbations. Part 1. Classical perturbations. Part 2. Quantum
  theory of perturbations. Part 3. Extensions},''
\href{http://dx.doi.org/10.1016/0370-1573(92)90044-Z}{{\em Phys. Rept.}
  {\bfseries 215} (1992) 203--333}.

\bibitem{Lyth:2009zz}
D.~H. Lyth and A.~R. Liddle, {\em {The primordial density perturbation:
  Cosmology, inflation and the origin of structure}}.
\newblock 2009.
\newblock
\url{http://www.cambridge.org/uk/catalogue/catalogue.asp?isbn=9780521828499}.
\newblock

\bibitem{Capozziello:2002rd}
S.~Capozziello, ``{Curvature quintessence},''
  \href{http://dx.doi.org/10.1142/S0218271802002025}{{\em Int. J. Mod. Phys.}
  {\bfseries D11} (2002) 483--492},
\href{http://arxiv.org/abs/gr-qc/0201033}{{\ttfamily arXiv:gr-qc/0201033
  [gr-qc]}}.

\bibitem{Amendola:2006kh}
L.~Amendola, D.~Polarski, and S.~Tsujikawa, ``{Are f(R) dark energy models
  cosmologically viable ?},''
  \href{http://dx.doi.org/10.1103/PhysRevLett.98.131302}{{\em Phys. Rev. Lett.}
  {\bfseries 98} (2007) 131302},
\href{http://arxiv.org/abs/astro-ph/0603703}{{\ttfamily arXiv:astro-ph/0603703
  [astro-ph]}}.

\bibitem{Nojiri:2006su}
S.~Nojiri and S.~D. Odintsov, ``{Modified gravity as an alternative for
  Lambda-CDM cosmology},''
  \href{http://dx.doi.org/10.1088/1751-8113/40/25/S17}{{\em J. Phys.}
  {\bfseries A40} (2007) 6725--6732},
\href{http://arxiv.org/abs/hep-th/0610164}{{\ttfamily arXiv:hep-th/0610164
  [hep-th]}}.

\bibitem{Starobinsky:2007hu}
A.~A. Starobinsky, ``{Disappearing cosmological constant in f(R) gravity},''
  \href{http://dx.doi.org/10.1134/S0021364007150027}{{\em JETP Lett.}
  {\bfseries 86} (2007) 157--163},
\href{http://arxiv.org/abs/0706.2041}{{\ttfamily arXiv:0706.2041 [astro-ph]}}.

\bibitem{Stelle:1976gc}
K.~Stelle, ``{Renormalization of Higher Derivative Quantum Gravity},''
\href{http://dx.doi.org/10.1103/PhysRevD.16.953}{{\em Phys.Rev.} {\bfseries
  D16} (1977) 953--969}.

\bibitem{weyl1918gravitation}
H.~Weyl, ``Gravitation and electricity,'' {\em Sitzungsber. K{\"o}nigl. Preuss.
  Akad. Wiss.} {\bfseries 26} (1918) 465--480.

\bibitem{buchdahl1962gravitational}
H.~Buchdahl, ``On the gravitational field equations arising from the square of
  the Gaussian curvature,'' {\em Il Nuovo Cimento Series 10} {\bfseries 23}
  no.~1, (1962) 141--157.

\bibitem{ruzmaikina1970quadratic}
T.~Ruzmaikina and A.~Ruzmaikin, ``Quadratic corrections to the Lagrangian
  density of the gravitational field and the singularity,'' {\em Sov. Phys.
  JETP} {\bfseries 30} (1970) 372.

\bibitem{Gurovich:1979xg}
V.~T. Gurovich and A.~A. Starobinsky, ``{Quantum effects and regular
  cosmological models},'' {\em Sov. Phys. JETP} {\bfseries 50} (1979) 844--852.
[Zh. Eksp. Teor. Fiz.77,1683(1979)].

\bibitem{tomita1978anisotropic}
K.~Tomita, T.~Azuma, and H.~Nariai, ``On anisotropic and homogeneous
  cosmological models in the renormalized theory of gravitation,'' {\em
  Progress of Theoretical Physics} {\bfseries 60} no.~2, (1978) 403--413.

\bibitem{Muller:1987hp}
V.~Muller, H.~Schmidt, and A.~A. Starobinsky, ``{The Stability of the De Sitter
  Space-time in Fourth Order Gravity},''
\href{http://dx.doi.org/10.1016/0370-2693(88)90007-X}{{\em Phys.Lett.}
  {\bfseries B202} (1988) 198}.

\bibitem{Berkin:1991nb}
A.~L. Berkin, ``{Contribution of the Weyl tensor to R**2 inflation},''
\href{http://dx.doi.org/10.1103/PhysRevD.44.1020}{{\em Phys.Rev.} {\bfseries
  D44} (1991) 1020--1027}.

\bibitem{Barrow:2005qv}
J.~D. Barrow and S.~Hervik, ``{Anisotropically inflating universes},''
  \href{http://dx.doi.org/10.1103/PhysRevD.73.023007}{{\em Phys.Rev.}
  {\bfseries D73} (2006) 023007},
\href{http://arxiv.org/abs/gr-qc/0511127}{{\ttfamily arXiv:gr-qc/0511127
  [gr-qc]}}.

\bibitem{vitenti2006numerical}
S.~D.~P. Vitenti and D.~M{\"u}ller, ``Numerical Bianchi type I solutions in
  semiclassical gravitation,'' {\em Physical Review D} {\bfseries 74} no.~6,
  (2006) 063508.

\bibitem{muller2006starobinsky}
D.~M{\"u}ller and S.~D.~P. Vitenti, ``About Starobinsky inflation,'' {\em
  Physical Review D} {\bfseries 74} no.~8, (2006) 083516.

\bibitem{cotsakis2008slice}
S.~Cotsakis, ``Slice energy in higher-order gravity theories and conformal
  transformations,'' {\em Gravitation and Cosmology} {\bfseries 14} no.~2,
  (2008) 176--183.

\bibitem{Barrow:2009gx}
J.~D. Barrow and S.~Hervik, ``{Simple Types of Anisotropic Inflation},''
  \href{http://dx.doi.org/10.1103/PhysRevD.81.023513}{{\em Phys.Rev.}
  {\bfseries D81} (2010) 023513},
\href{http://arxiv.org/abs/0911.3805}{{\ttfamily arXiv:0911.3805 [gr-qc]}}.

\bibitem{muller2011homogeneous}
D.~M{\"u}ller, ``Homogeneous solutions of quadratic gravity,'' in {\em
  International Journal of Modern Physics: Conference Series}, vol.~3,
  pp.~111--120, World Scientific.
\newblock 2011.

\bibitem{de2012bianchi}
J.~A. de~Deus and D.~M{\"u}ller, ``Bianchi VII A solutions of effective
  quadratic gravity,'' {\em General Relativity and Gravitation} {\bfseries 44}
  no.~6, (2012) 1459--1478.

\bibitem{Muller:2012wx}
D.~Muller and J.~A. de~Deus, ``{Bianchi I solutions of effective quadratic
  gravity},'' \href{http://dx.doi.org/10.1142/S021827181250037X}{{\em
  Int.J.Mod.Phys.} {\bfseries D21} (2012) 1250037},
\href{http://arxiv.org/abs/1203.6882}{{\ttfamily arXiv:1203.6882 [gr-qc]}}.

\bibitem{MULLER:2014jaa}
D.~M\"uller, M.~E. Alves, and J.~C. de~Araujo, ``{The Isotropization Process in
  the Quadratic Gravity},''
\href{http://dx.doi.org/10.1142/S0218271814500199}{{\em Int.J.Mod.Phys.}
  {\bfseries D23} (2014) 1450019}.

\bibitem{0264-9381-27-22-225013}
J.~Middleton, ``On the existence of anisotropic cosmological models in higher
  order theories of gravity,'' {\em Classical and Quantum Gravity} {\bfseries
  27} no.~22, (2010) 225013.
  \url{http://stacks.iop.org/0264-9381/27/i=22/a=225013}.

\bibitem{PhysRevD.77.103523}
J.~Middleton and J.~D. Barrow, ``Stability of an isotropic cosmological
  singularity in higher-order gravity,''
  \href{http://dx.doi.org/10.1103/PhysRevD.77.103523}{{\em Phys. Rev. D}
  {\bfseries 77} (May, 2008) 103523}.
  \url{http://link.aps.org/doi/10.1103/PhysRevD.77.103523}.

\bibitem{PhysRevD.75.123515}
J.~D. Barrow and J.~Middleton, ``Stable isotropic cosmological singularities in
  quadratic gravity,'' \href{http://dx.doi.org/10.1103/PhysRevD.75.123515}{{\em
  Phys. Rev. D} {\bfseries 75} (Jun, 2007) 123515}.
  \url{http://link.aps.org/doi/10.1103/PhysRevD.75.123515}.

\bibitem{Cotsakis:2007un}
S.~Cotsakis and A.~Tsokaros, ``{Asymptotics of flat, radiation universes in
  quadratic gravity},''
  \href{http://dx.doi.org/10.1016/j.physletb.2007.06.038}{{\em Phys. Lett.}
  {\bfseries B651} (2007) 341--344},
\href{http://arxiv.org/abs/gr-qc/0703043}{{\ttfamily arXiv:gr-qc/0703043
  [GR-QC]}}.

\bibitem{Cotsakis:1997ck}
S.~Cotsakis and J.~Miritzis, ``{Proof of the cosmic no hair conjecture for
  quadratic homogeneous cosmologies},''
  \href{http://dx.doi.org/10.1088/0264-9381/15/9/024}{{\em Class. Quant. Grav.}
  {\bfseries 15} (1998) 2795--2801},
\href{http://arxiv.org/abs/gr-qc/9712026}{{\ttfamily arXiv:gr-qc/9712026
  [gr-qc]}}.

\bibitem{Miritzis:2003eu}
J.~Miritzis, ``{Dynamical system approach to FRW models in higher order gravity
  theories},'' \href{http://dx.doi.org/10.1063/1.1602161}{{\em J. Math. Phys.}
  {\bfseries 44} (2003) 3900--3910},
\href{http://arxiv.org/abs/gr-qc/0305062}{{\ttfamily arXiv:gr-qc/0305062
  [gr-qc]}}.

\bibitem{Miritzis:2007yn}
J.~Miritzis, ``{Oscillatory behavior of closed isotropic models in second order
  gravity theory},'' \href{http://dx.doi.org/10.1007/s10714-008-0651-3}{{\em
  Gen. Rel. Grav.} {\bfseries 41} (2009) 49--65},
\href{http://arxiv.org/abs/0708.1396}{{\ttfamily arXiv:0708.1396 [gr-qc]}}.

\bibitem{Muller:1989rp}
V.~Muller, H.~J. Schmidt, and A.~A. Starobinsky, ``{Power law inflation as an
  attractor solution for inhomogeneous cosmological models},''
\href{http://dx.doi.org/10.1088/0264-9381/7/7/012}{{\em Class. Quant. Grav.}
  {\bfseries 7} (1990) 1163--1168}.

\bibitem{Barrow:2006xb}
J.~D. Barrow and S.~Hervik, ``{On the evolution of universes in quadratic
  theories of gravity},''
  \href{http://dx.doi.org/10.1103/PhysRevD.74.124017}{{\em Phys.Rev.}
  {\bfseries D74} (2006) 124017},
\href{http://arxiv.org/abs/gr-qc/0610013}{{\ttfamily arXiv:gr-qc/0610013
  [gr-qc]}}.

\bibitem{Kasner:1921zz}
E.~Kasner, ``{Geometrical theorems on Einstein's cosmological equations},''
\href{http://dx.doi.org/10.2307/2370192}{{\em Am.J.Math.} {\bfseries 43} (1921)
  217--221}.

\bibitem{Toporensky:2016kss}
A.~Toporensky and D.~M{\"u}ller, ``{On stability of the Kasner solution in
  quadratic gravity},'' \href{http://dx.doi.org/10.1007/s10714-016-2172-9}{{\em
  Gen. Rel. Grav.} {\bfseries 49} no.~1, (2017) 8},
\href{http://arxiv.org/abs/1603.02851}{{\ttfamily arXiv:1603.02851 [gr-qc]}}.

\bibitem{wainwright2005dynamical}
J.~Wainwright and G.~F.~R. Ellis, {\em Dynamical systems in cosmology}.
\newblock Cambridge University Press, 2005.

\bibitem{Netto:2015cba}
T.~d.~P. Netto, A.~M. Pelinson, I.~L. Shapiro, and A.~A. Starobinsky, ``{From
  stable to unstable anomaly-induced inflation},''
  \href{http://dx.doi.org/10.1140/epjc/s10052-016-4390-4}{{\em Eur. Phys. J.}
  {\bfseries C76} no.~10, (2016) 544},
\href{http://arxiv.org/abs/1509.08882}{{\ttfamily arXiv:1509.08882 [hep-th]}}.

\bibitem{Kamenshchik:2017fk}
A.~Kamenshchik, E.~Pozdeeva, A.~Starobinsky, A.~Tronconi, G.~Venturi, and
  S.~Vernov, ``Induced gravity, and minimally and conformally coupled scalar
  fields in Bianchi-I cosmological models,''
  \href{http://arxiv.org/abs/1710.02681}{{\ttfamily 1710.02681}}.
  \url{https://arxiv.org/abs/1710.02681}.

\bibitem{Sotiriou:2008rp}
T.~P. Sotiriou and V.~Faraoni, ``{f(R) Theories Of Gravity},''
  \href{http://dx.doi.org/10.1103/RevModPhys.82.451}{{\em Rev. Mod. Phys.}
  {\bfseries 82} (2010) 451--497},
\href{http://arxiv.org/abs/0805.1726}{{\ttfamily arXiv:0805.1726 [gr-qc]}}.

\bibitem{Belinsky:1970ew}
V.~Belinsky, I.~Khalatnikov, and E.~Lifshitz, ``{Oscillatory approach to a
  singular point in the relativistic cosmology},''
\href{http://dx.doi.org/10.1080/00018737000101171}{{\em Adv.Phys.} {\bfseries
  19} (1970) 525--573}.

\end{thebibliography}\endgroup


\end{document}